\documentclass[usegraphicx]{mn2e}

\usepackage{mathptmx}

\newcommand{\Mpc}{\rm\thinspace Mpc}

\newcommand{\km}{\rm\thinspace km}

\newcommand{\yr}{\rm\thinspace yr}

\newcommand{\s}{\rm\thinspace s}

\newcommand{\kmps}{\hbox{$\km\s^{-1}\,$}}

\newcommand{\kmpspMpc}{\hbox{$\kmps\Mpc^{-1}$}}

\begin{document}
\title{Isothermal shocks in Abell 2199 and 2A~0335+096?}

\author[J.S. Sanders and A.C. Fabian]{J.S.  Sanders\thanks{E-mail:
    jss@ast.cam.ac.uk} and
  A.C. Fabian\\
  Institute of Astronomy, Madingley Road, Cambridge. CB3 0HA}
\maketitle

\begin{abstract}
  We report on a partially circular X-ray surface brightness
  discontinuity found at about 55~kpc from the centre of Abell~2199
  with \emph{Chandra X-ray observatory} observations. Unlike cold
  fronts found in other clusters, the feature shows no significant
  temperature change across it but has an apparent density jump. We
  therefore identify it as a weak isothermal shock associated with the
  central AGN and the inflation of its radio bubbles, as found in the
  Perseus cluster.  We examine a similar feature at 40~kpc radius
  found by Mazzotta et al in 2A~0335+096, and conclude that it too may
  be an isothermal shock.  The change in density if these are shocks
  implies a Mach number of $\sim 1.5$.  If the isothermal nature of
  these features is confirmed by deeper observations, the implication
  is that such shocks are common in clusters of galaxies, and are an
  important mechanism for the transport of energy from a central
  supermassive black hole into the cluster core.
\end{abstract}


\section{Introduction}
Abell~2199 is a regular, rich, X-ray bright cluster of galaxies at a
redshift of 0.0309.  The X-ray emission is steeply peaked (Peres et al
2001). \emph{Chandra} X-ray observations of the cluster show that it
is regular near the centre, although there is more bright X-ray
emission to the south where the older parts of the radio source lie
(Johnstone et al 2002).  Abell~2199 hosts the unusual radio galaxy
3C338 which shows evidence for multiple episodes of radio emission.
The extended, older emission clearly coincides with two outer large
depressions 25~kpc either side of the nucleus.

2A~0335+096 is another bright, nearby ($z=0.0349$) galaxy cluster.
\emph{Chandra} observations of the cluster found complex substructure
in the core (Kawano, Ohto \& Fukazawa 2003; Mazzotta, Edge \&
Markevitch 2003), consisting of a number a X-ray bright blobs and a
``cold front'' 40~kpc to the south of the centre. The cold front is
peculiar, showing no temperature discontinuity across it (Mazzotta et
al 2003). The cluster hosts another unusual radio source (Sarazin et
al 1995), and was the subject of a deep \emph{XMM-Newton} observation
(Werner et al 2006).

In this letter we examine \emph{Chandra} observations of Abell~2199
and 2A~0335+096. We have discovered a jump in surface brightness in
Abell~2199 without evidence for any temperature change across it. We
identify this as an isothermal shock. We also investigate the ``cold
front'' in 2A~0335+096, and conclude it is an isothermal shock
too. Finally we examine a \emph{Chandra} image of H~1821+643
($z=0.297$) and report a suggestive ring structure.

We assume $H_0 = 70 \kmpspMpc$ when calculating distances, so the
angular scales are 0.61, 0.685 and 4.39 kpc arcsec${}^{-1}$ for
Abell~2199, 2A~0335+096 and H~1821+643, respectively.

\section{Data analysis}
\subsection{Abell 2199}
The analysis was undertaken with \emph{Chandra} OBSIDs 498 and 497,
taken with the detectors at $-110^\circ$C and $-120^\circ$C,
respectively. These datasets used the ACIS-S array for imaging. The
event files were reprocessed with \textsc{ciao} 3.3.0.1, using
\textsc{caldb} 3.10. Light curves for the event files were extracted
from the ACIS-S1 CCD in the 2.5 to 7~keV band to filter the data for
flares. This CCD uses the same front-illuminated technology as the
ACIS-S3 CCD, and so is a good choice for filtering for flares. This
energy band is optimised for flare filtering (Markevitch 2002). We
filtered out flared time periods by eye from each dataset. This
yielded a total of 33.1~ks clean exposure.

For spectral fitting, background spectra were extracted from blank-sky
fields, reprocessed and reprojected to match their respective
foreground observations. The exposure of the background observation
was adjusted to give the same count rate in the 10-12~keV band as the
foreground observation.  Response matrices were created using the
\textsc{ciao} \textsc{mkrmf} tool, weighting according to the number
of counts between 0.5 and 7~keV in each spectral region, and ancillary
response matrices were made using \textsc{mkwarf}.

\subsection{2A~0335+096}
For 2A~0335+096, we analysed the \emph{Chandra} OBSID 919. This
observation again used the ACIS-S3 CCD for imaging. We reprocessed the
data in the same way as above. As was noted by Mazzotta et al (2003),
most of this observation is affected by a mild flare. Using the
ACIS-S1 CCD, we removed periods where the count rate was above 0.35
counts per second in the above band. This yielded a total exposure of
11.8~ks. Using a slightly less strict time selection, Mazzotta et al
(2003) noted that the cluster emission is dominant over the flare, and
the inclusion of a component for the flare in the modelling makes very
little difference to the results. We therefore ignore the flare in our
modelling. For this observation we again made background spectra from
a blank-sky observation.  Responses were created using
\textsc{mkacisrmf}.

\section{Results}

\begin{figure*}
  \centering
  \includegraphics[width=0.6\textwidth]{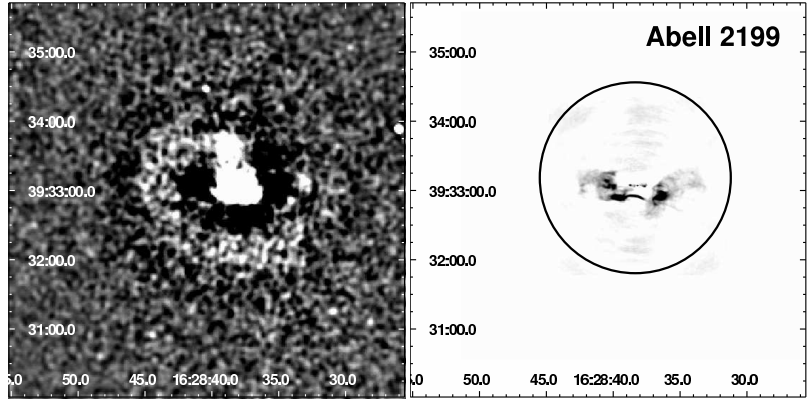}
  \caption{(Left) Unsharp-masked image of the core of Abell~2199 in
    the 0.5 to 6~keV band.  (Right) 1.4~GHz VLA radio map of the same
    region (Giovannini et al 1998). The position of the X-ray edge is
    marked by a circle.}
  \label{fig:a2199_unsharp}
\end{figure*}

In Fig.~\ref{fig:a2199_unsharp} we show an unsharp-masked image of
Abell~2199. An exposure-map corrected image of the cluster was made
using both datasets in the 0.5 to 6~keV band. This image was smoothed
by Gaussians of 1.96 and 19.6~arcsec, and the second image was
subtracted from the first. This procedure enhances features between
these two scales, by subtracting the variation on larger and smaller
scales. Immediately a semi-circular ring shaped feature to the south
and east of the core.  We also plot on the figure the 1.4~GHz radio
emission. As reported by Johnstone et al (2002), the radio emission
occupies holes in the X-ray emission. There are additional ``fossil''
radio bubbles which have no radio association, found to the south of
the nucleus.

\begin{figure*}
  \centering
  \includegraphics[width=0.6\textwidth]{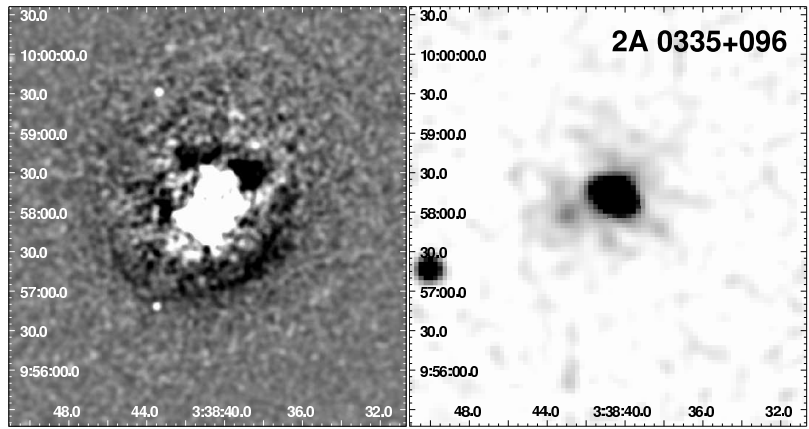}
  \caption{(Left) Unsharp-masked image of the core of 2A~0335+096,
    using the same smoothing parameters as
    Fig.~\ref{fig:a2199_unsharp}.  (Right) C-band VLA radio map of the
    same region.}
  \label{fig:2a0335_unsharp}
\end{figure*}

2A~0335+096 shows a similar surface brightness discontinuity to the
south of its core (Fig.~\ref{fig:2a0335_unsharp}), as was noted by
Mazzotta et al (2003).

\begin{figure*}
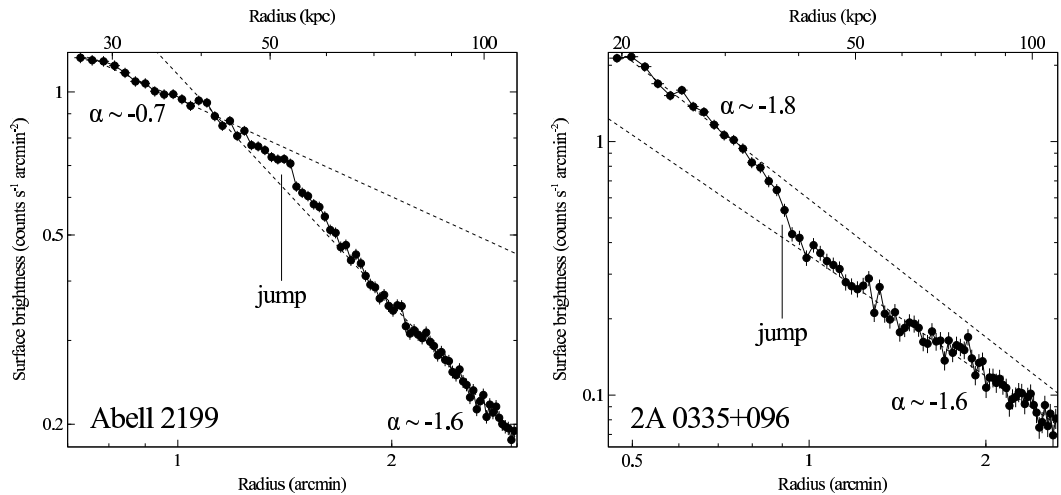

  \centering
  \includegraphics[width=0.8\columnwidth]{fix_a2199_SB.epsi}
  \hspace{3mm}
  \includegraphics[width=0.8\columnwidth]{fix_2a0335_SB.epsi}
  \caption{(Left) Surface brightness profile across the discontinuity
    in Abell~2199 in the 0.5 to 6~keV band. Note the change in slope
    around 1.4~arcmin. The profile was made between angles 186 and 290
    degrees from the north from the west, and was centred on the
    position $16^{h}28^{m}38.0^{s}$, $+39^\circ 33' \: 24''$ (J2000).
    (Right) Surface brightness profile for 2A~0335+096. This was made
    between angles 246 and 351 degrees, and centred on
    $03^{h}38^{m}41.3^{s}$, $+09^\circ 57' \: 57''$. The plots show
    powerlaw fits to the inner and outer parts of the profiles.}
  \label{fig:sb}
\end{figure*}

We have created surface brightness profiles across these features
(Fig.~\ref{fig:sb}), where they are most circular in shape on the sky.
In the central region of Abell~2199, the surface brightness is
relatively flat. A powerlaw fit gives a slope of around -0.7. The
surface brightness profile then quickly steepens, giving a slope of
-1.6 at large radii. In 2A~0335+096, the central surface profile is
steeper than Abell~2199, with an index of -1.8. There appears to be a
surface brightness discontinuity at a radius of around 0.9~arcmin. The
surface brightness then appears to decline with a similar slope to the
interior, with an index of around -1.6.

\begin{figure*}
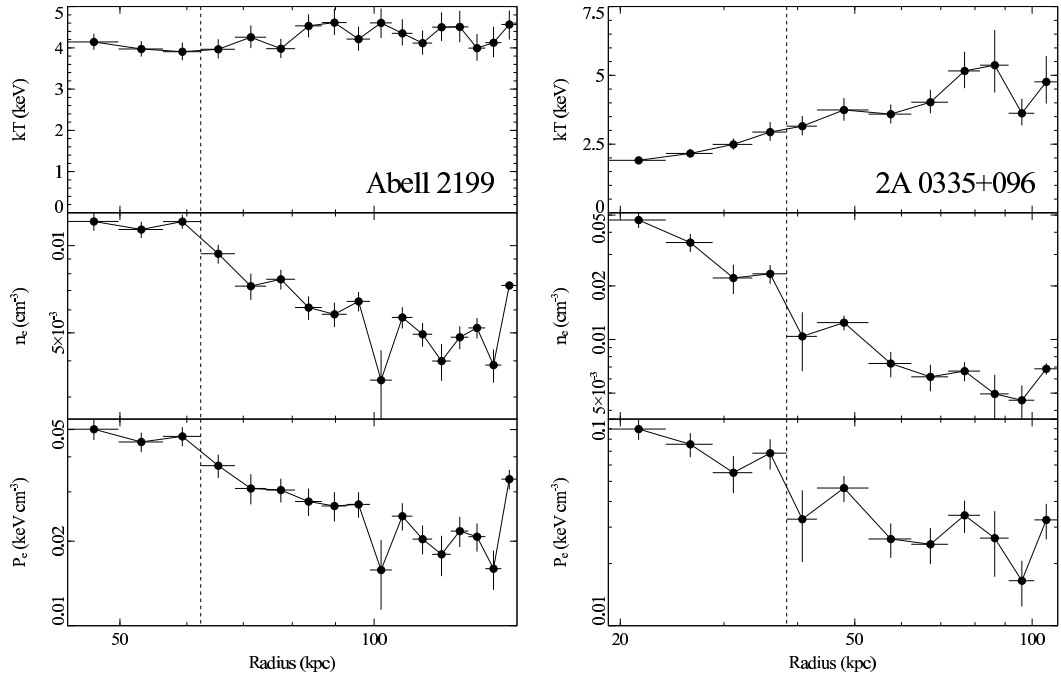

  \centering
  \includegraphics[width=0.8\columnwidth]{fix_a2199_prof.epsi}
  \hspace{3mm}
  \includegraphics[width=0.8\columnwidth]{fix_2a0335_prof.epsi}
  \caption{Projected temperature, deprojected density, and pressure
    across the surface brightness jump in Abell~2199 (left) and
    2A~0335+096 (right). The dotted lines show the approximate
    positions of the surface brightness discontinuities.}
  \label{fig:profs}
\end{figure*}

We have also measured temperature profiles across these features.
Using sectors with the same range of angles as the surface brightness
profiles, we fitted single temperature \textsc{mekal} models to
spectra extracted from annuli. In the fits the absorption,
temperature, metallicity relative to solar, and normalisation were
free. These resulting temperature profiles are plotted in
Fig.~\ref{fig:profs}. Also shown are dotted lines at the approximate
radii of the discontinuities. From the emission measures we calculated
deprojected electron densities in each shell, assuming spherical
symmetry. Working from the outer annulus, the contribution to the
emission measure from annuli external to the one being examined were
subtracted, to calculate the density. Approximate pressures were
calculated by multiplying the projected temperature and deprojected
density.

The deprojected densities we calculated are very similar to the values
produced using the \textsc{projct} model in \textsc{xspec}, which
accounts for projection. The projected temperatures show very little
difference to the deprojected ones from \textsc{projct}, but have
smaller error bars. We do not find any evidence for metallicity jumps
across the surface brightness changes, although the uncertainty on the
abundance measurements could hide moderate changes.

\section{Discussion}
It can be seen that the density abruptly declines outward across each
of these surface brightness features. In Abell~2199 the density drops
by around 40~per~cent, and in 2A~0335+096 it falls by 45~per~cent.
However there is no significant temperature change over either edge.
This means that there are discontinuities in the pressure profiles,
implying that these features are not static. The statistics do not
rule out the presence of small temperature jumps, but if there are
none then they are not cold-fronts (Markevitch et al 2000), which are
continuous in pressure.

We therefore identify the features are isothermal shocks, as
discovered in the Perseus cluster (Fabian et al 2006).  Shocks may
appear to be isothermal if conduction is operating efficiently. As the
electrons move faster than the ions, they can travel ahead of the
shock. This means that the increase in electron temperature (which we
measure from the X-ray spectra) is smoothed out over a large region,
while the ion temperature is discontinuous over the shock.

Using equation (10-2) of Spitzer (1978), for a non-radiating shock the
pre and postshock densities, $\rho_1$ and $\rho_2$, are related to
the Mach number of shock, $M$, by
\begin{equation}
  \frac{\rho_1}{\rho_2} = \frac{\gamma - 1}{\gamma + 1} +
  \frac{2}{(\gamma+1) M^2}.
\end{equation}
Using a value of the ratio of the specific heats, $\gamma$, of 5/3,
this yields a Mach number, $M$, of around 1.5 for both shocks. If the
shocks were not isothermal, this would imply a 50~per~cent rises in
temperature which are not observed.

\begin{figure}
  \centering
  \includegraphics[width=0.8\columnwidth]{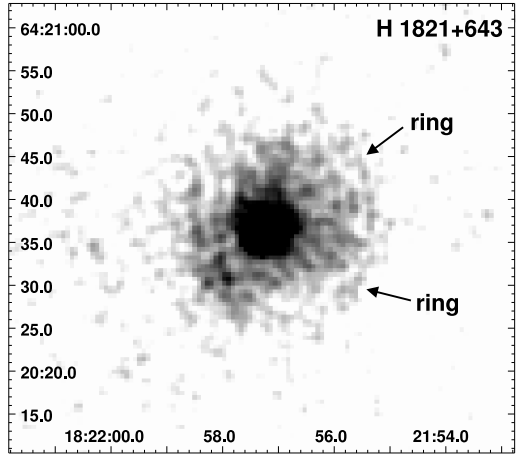}
  \caption{99.6~ks image of the central region of the cluster around
    H~1821+643 in the 0.5 to 3~keV band, smoothed with a Gaussian of
    0.5~arcsec. This image is the zeroth order LETG image. The
    dispersion directions lie along SW-NE and SWW-NEE directions. This
    is \emph{Chandra} OBS\_ID 1599.}
  \label{fig:h1821}
\end{figure}

A similar shock may exist in the cluster surrounding the quasar
H~1821+643. In Fig.~\ref{fig:h1821} we show a 99.6~ks \emph{Chandra}
Low Energy Transmission Grating (LETG) zeroth order image of the
cluster. The effective area of the instrument with the the LETG in
place is reduced by a factor of 6-8, so the image represents a
partial, noisy, snapshot. Visible in the image is a ring around the
core, roughly 10~arcsec (44~kpc) in radius. Also seen is a 5~arcsec
long mushroom shaped feature to the north of the nucleus, which is
spatially coincident with, and has the same morphology as its radio
lobe.

As we have found isothermal shocks in three nearby cool core clusters,
it appears likely that they are widespread. If this is the case, they
are likely to be an important mechanism for the transport of energy
from the central active nucleus to the surrounding cluster. The
discovery of an isothermal shock in Perseus was completely unexpected,
but they could be very important for unravelling the heating and
cooling processes in clusters. In addition, their existence may
indicate that the magnetic field in cluster cores may be radial, for
conduction to be operating efficiently across shocks. However not all
shocks are isothermal, for example the weak shocks in Cygnus A (Wilson,
Smith \& Young 2006) and M87 (Forman et al 2006).

We note that there is a problem for the clusters if the shocks are not
isothermal, since the temperature jump should be much higher at
smaller radii. The density profile of the cool cores of clusters
approximately varies inversely with radius ($n\propto r^{-1}$), so a
pressure driven shock has a velocity $v$ which evolves with radius $r$
as $v\propto r^{-1}(E/n)^{1/2}$. A jump of say 0.5~keV at radius
50~kpc would have had a shock temperature ($T\propto v^2$) of over
10~keV when at 10~kpc, which is around the size of the bubbles seen.
This problem is particularly acute for the M87 shock where Forman et
al (2006) report a jump of 0.5~keV at 13~kpc when the typical bubble
size is 1.5~kpc. The temperature of the gas is observed to drop inward
whereas a shock would predict much higher inner temperatures, even
though the gas would cool somewhat by adiabatic expansion after the
shock has passed.

Repeat efficient shocking of gas in cool cluster cores would soon
create temperature, and entropy, profiles disagreeing markedly from
the steady, inward drop, generally found (at least one third of clusters
have such cool cores, Dunn et al 2005). 

We have investigated several processes which could tend to make the
gas appear isothermal. These include electron-proton coupling, and
non-ionization equilibrium. The latter has too short a timescale (less
than one million yr) to be relevant, whereas the former takes about
$2\times 10^7\yr$. However compression of both electrons and protons
dominates the heating in weak shocks so it is unclear whether it would
have a significant effect. The conduction timescale is about $5\times
10^6 \ell^2_1\yr,$ where $\ell_1$ is the distance from the shock in
units of 10~kpc and Spitzer electron conductivity is assumed. The
shock propagation timescale is about 10 Myr per 10~kpc, so if
conduction is efficient then it can operate to make the shock appear
isothermal. The conclusion on the nature of the shocks is therefore
similar to that reached for the Perseus cluster (Fabian et al 2006). 

Cosmic rays can also pass through the shock front and so transfer
energy from shocked to unshocked gas. The inner regions of cool cores
are likely to contain a significant cosmic-ray population (see e.g.
the discussion of the Perseus radio mini-halo in Sanders et al 2005).
If much of the shock energy passes into the cosmic-rays and magnetic
field, rather than into the thermal X-ray emitting population, then
the thermal gas can appear to be isothermal. We speculate that the
cosmic rays can then release much of the shock energy at later times
and over a wider region. This possibility may vary from cluster to
cluster depending on the energy density of the cosmic rays in the
relevant region.

An alternative possibility is that some features are due to changes in
metallicity, rather like the metal-rich ring found in the Perseus
cluster (Sanders et al 2005). As reported at the end of Section 3, we
find no metallicity jumps which could explain the features on
Abell~2199 or 2A\,0335+096.

Deeper X-ray observations of these cluster cores are required to
understand the nature of the shock-like features further and rule out
the presence of any cold front. Both A2199 and 2A\,0335+096 have X-ray
bright cores (0.7 and 0.5 ct~s$^{-1}$~arcmin$^{-2}$, respectively at
centre of the jump) and relatively brief current Chandra observations.
Exposures which are 5--10 times deeper are feasible and would enable
the shape and temperatures of the features to be characterized in
detail.

\section*{Acknowledgements}
ACF thanks the Royal Society for support.

\clearpage
\end{document}